\documentclass[english]{elsarticle}
\usepackage[T1]{fontenc}
\usepackage[latin9]{inputenc}
\usepackage[a4paper]{geometry}
\usepackage{array}
\usepackage{amsmath}
\usepackage{graphicx}

\makeatletter

\providecommand{\tabularnewline}{\\}

\@ifundefined{showcaptionsetup}{}{%
 \PassOptionsToPackage{caption=false}{subfig}}
\usepackage{subfig}
\makeatother

\usepackage{babel}

\begin{document}

\title{Complex networks derived from cellular automata}

\author{Yoshihiko Kayama}

\ead{kayama@baika.ac.jp}

\address{Department of Media and Information, BAIKA Women's University, \\
2-19-5, Shukuno-sho, Ibaraki-city, Osaka-pref., Japan}
\begin{abstract}
We propose a method for deriving networks from one-dimensional binary
cellular automata. The derived networks are usually directed and have
structural properties corresponding to the dynamical behaviors of
their cellular automata. Network parameters, particularly the efficiency
and the degree distribution, show that the dependence of efficiency
on the grid size is characteristic and can be used to classify cellular
automata and that derived networks exhibit various degree distributions.
In particular, a class IV rule of Wolfram's classification produces
a network having a scale-free distribution.\end{abstract}
\begin{keyword}
complex networks, cellular automata, scale-free, small-world
\end{keyword}
\maketitle

\section{Introduction}

Cellular automata (CA) have been used to study the critical phenomena
of complex systems. S. Wolfram has systematically investigated the
dynamical behavior of one-dimensional automata and identified the
following four essential types: homogeneous (class I), periodic (class
II), chaotic (class III), and complex (class IV) \citep{Wolfram1983}.
In particular, class IV rules produce complex structures with long
transients. Alternatively, complex networks exhibiting a scale-free
topology can be identified ubiquitously; for example, they occur in
social relationships \citep{Wasserman1994}, biological and chemical
systems \citep{Jeong2000,Jeong2001}, and the Internet \citep{R.Albert1999,Barabasi2000}.
Such networks have been studied extensively in the wake of papers
by Watts and Strogatz on small-world networks \citep{D.J.Watts1998}
and by Barabási and Albert on scale-free networks \citep{Barabasi1999}.
Even though both CA and complex networks are used to study complex
systems and various phenomena, their relation is not clear.

In recent years, the dynamics of Boolean networks with complex topology
have been studied \citep{O'Sullivan2001}-\nocite{Aldana2003,Aldana2003a,Darabos2007}\citep{Marr2009}
yielding several notable results such as the dependence of dynamical
phases on network topology \citep{Aldana2003} and the robustness
of scale-free networks \citep{Aldana2003a}. Our approach is to define
networks that correspond to the dynamical behaviors of CA.

In this article, we propose a method for deriving networks from the
time evolution of CA configurations. For transformations of the CA
rule function, the adjacency matrix of the network exhibits characteristic
properties. We investigate networks derived from the typical rules
of one-dimensional binary CA with three and five neighbors. Our studies
reveal that the \textit{structural} properties of the derived network
reflect the \emph{dynamical} behaviors of the CA rule and that chaotic
or critical rules lead to complex network topologies. We use two parameters
to characterize the topology of such networks: the efficiency of the
network (corresponding to the harmonic mean of its shortest path lengths)
and its degree distribution. The efficiency shows a characteristic
scale dependence on the grid size of the cellular automaton and may
be useful in classifying CA rules. Class III rules correspond to random
networks, and a scale-free degree distribution has been obtained from
the class IV rule.

The next section describes our notation and some definitions relevant
to CA. In section 3, we describe our method for deriving networks
from CA rules and discuss their properties under transformations of
the CA rule function. Section 4 reports on the efficiency and degree
distribution of derived networks and discusses the correspondence
between the dynamical behavior of a cellular automaton and the structural
properties of its derived network.

\section{Notation and definitions relevant to cellular automata}

CA are dynamical systems that consist of a regular grid of cells,
each characterized by a finite number of states. CA are updated synchronously
in discrete time steps according to a local rule (CA rule) that is
identical at every cell. In a one-dimensional grid, each cell is connected
to its $r$ local neighbors on either side, where $r$ is referred
to as the \textit{radius.} Thus, each cell has $2r+1$ neighbors,
including itself. The state of a cell for the next time step is determined
from the current states of the neighboring cells: $x_{i}(t+1)=f(x_{i-r}(t),...,x_{i}(t),...,x_{i+r}(t))$,
where $x_{i}(t)$ denotes the state of cell $i$ at time $t$, and
$f$ is the transition rule function. The term \textit{configuration}
refers to an assignment of states to all the cells for a given time
step; a configuration is denoted by $\boldsymbol{x}(t)=(x_{0}(t),x_{1}(t),...,x_{N-1}(t))$,
where \textit{N} is the grid size. Thus, the time transition of a
configuration $\boldsymbol{x}(t)$ with periodic boundary conditions
is given by

\begin{eqnarray}
\boldsymbol{x}(t+1) & = & \boldsymbol{f}(\boldsymbol{x}(t))\\
 & = & (f(x_{N-r}(t),\ldots,x_{0}(t),\ldots,x_{r}(t)),\nonumber \\
 &  & \quad f(x_{N-r+1}(t),\ldots,x_{1}(t),\ldots,x_{1+r}(t)),\nonumber \\
 &  & \quad\ldots,f(x_{N-1-r}(t),\ldots,x_{N-1}(t),\ldots,x_{r-1}(t))),\end{eqnarray}

\begin{flushleft}
where $\boldsymbol{f}$ represents a mapping on the configuration
space $\left\{ \boldsymbol{x}\right\} _{N}$. In this article, we
restrict our discussion to\emph{ binary} CA, which satisfy $x_{i}\in\left\{ 0,1\right\} $
for all \emph{i}.
\par\end{flushleft}

For a given configuration $\boldsymbol{x}$, the mirror (left-right
reflection) and complement (0-1 exchange) configurations are denoted
as $\tilde{\boldsymbol{x}}\equiv(x_{N-1},...,x_{1},x_{0})$ and $\boldsymbol{\bar{x}}\equiv(\bar{x}_{0},\bar{x}_{1},...,\bar{x}_{N-1})$,
respectively. By analogy, the mirror, complement and mirror-complement
of a rule function $f$ are defined as follows:

\begin{eqnarray}
\tilde{f}(x_{i-r},...,x_{i},...,x_{i+r}) & \equiv & f(x_{i+r},...,x_{i},...,x_{i-r})\label{eq:LR}\\
\bar{f}(x_{i-r},...,x_{i},...,x_{i+r}) & \equiv & \overline{f(\bar{x}_{i-r},...,\bar{x}_{i},...,\bar{x}_{i+r})}\label{eq:01}\\
\bar{\tilde{f}}(x_{i-r},...,x_{i},...,x_{i+r}) & \equiv & \overline{f(\bar{x}_{i+r},...,\bar{x}_{i},...,\bar{x}_{i-r})},\label{eq:LR-01}\end{eqnarray}

\begin{flushleft}
respectively. The CA rules of these transformed functions are equivalent
to the original rule $f$ \citep{Li1990}. It is trivial that mirror
and complement operations are commutative, i.e. $\bar{\tilde{f}}=\tilde{\bar{f}}$.
The mappings defined from these functions are \begin{eqnarray}
\tilde{\boldsymbol{f}}(\boldsymbol{x}) & \equiv & (\tilde{f}(x_{N-r},\ldots,x_{0},\ldots,x_{r}),\tilde{f}(x_{N-r+1},\ldots,x_{1},\ldots,x_{1+r}),\nonumber \\
 &  & \quad\ldots,\tilde{f}(x_{N-1-r},\ldots,x_{N-1},\ldots,x_{r-1}))\\
 & = & (f(x_{r},\ldots,x_{0},\ldots,x_{N-r}),f(x_{1+r},\ldots,x_{1},\ldots,x_{N-r+1}),\nonumber \\
 &  & \quad\ldots,f(x_{r-1},\ldots,x_{N-1},\ldots,x_{N-1-r}))\\
 & = & \widetilde{\boldsymbol{f}(\tilde{\boldsymbol{x}})}\label{eq:LtoR}\\
\bar{\boldsymbol{f}}(\boldsymbol{x}) & \equiv & (\bar{f}(x_{N-r},\ldots,x_{0},\ldots,x_{r}),\bar{f}(x_{N-r+1},\ldots,x_{1},\ldots,x_{1+r}),\nonumber \\
 &  & \quad\ldots,\bar{f}(x_{N-1-r},\ldots,x_{N-1},\ldots,x_{r-1}))\\
 & = & (\overline{f(\bar{x}_{N-r},\ldots,\bar{x}_{0},\ldots,\bar{x}_{r})},\overline{f(\bar{x}_{N-r+1},\ldots,\bar{x}_{1},\ldots,\bar{x}_{1+r})},\nonumber \\
 &  & \quad\ldots,\overline{f(\bar{x}_{N-1-r},\ldots,\bar{x}_{N-1},\ldots,\bar{x}_{r-1})})\\
 & = & \overline{\boldsymbol{f}(\boldsymbol{\bar{x}})}\label{eq:0to1}\\
\bar{\tilde{\boldsymbol{f}}}(\boldsymbol{x}) & = & \overline{\widetilde{\boldsymbol{f}(\bar{\tilde{\boldsymbol{x}}})}}=\tilde{\bar{\boldsymbol{f}}}(\boldsymbol{x}).\label{eq:LR01-01LR-f}\end{eqnarray}

\par\end{flushleft}

\begin{flushleft}
A \textit{t-}fold repetition of these mappings yields, respectively,
\par\end{flushleft}

\begin{eqnarray}
\tilde{\boldsymbol{f}}^{t}(\boldsymbol{x}) & = & \widetilde{\boldsymbol{f}_{R}^{t}(\tilde{\boldsymbol{x}})}\label{eq:LR-t-f}\\
\bar{\boldsymbol{f}}^{t}(\boldsymbol{x}) & = & \overline{\boldsymbol{f}_{R}^{t}(\bar{\boldsymbol{x}})}\label{eq:01-t-f}\\
\bar{\tilde{\boldsymbol{f}}}^{t}(\boldsymbol{x}) & = & \overline{\widetilde{\boldsymbol{f}_{R}^{t}(\bar{\tilde{\boldsymbol{x}}})}}=\tilde{\bar{\boldsymbol{f}}}^{t}(\boldsymbol{x}).\label{eq:LR01-01LR-t-f}\end{eqnarray}

\textit{Elementary Cellular Automata} (ECA) are the simplest nontrivial
binary CA; they are defined on a one-dimensional grid with minimal
neighborhood size ($r=1$). The $2^{3}=8$ different neighborhood
configurations result in $2^{8}=256$ possible rules, of which 88
are nonequivalent under the transformations \eqref{eq:LR}-\eqref{eq:LR-01}
\citep{Li1990}. ECA rules are generally referred to by their Wolfram
code, a standard naming convention invented by Wolfram \citep{Wolfram1983,Wolfram2002}
that gives each rule a number from 0 to 255. For example, rule 30
exhibits class III behavior, meaning that even simple input patterns
lead to chaotic, seemingly random histories. Rule 90 is also chaotic
with a fractal structure (Sierpinski triangle). Furthermore, rule
110 generates class IV behavior, which is neither completely random
nor completely repetitive. Localized structures appear and interact
in various complicated ways. 

Other simple models are 5-neighbor $(r=2)$ CA, which contain $2^{32}$
rules. Our discussion is restricted to the \textit{totalistic} CA
(5TCA), in which the state of each cell at time $t$ depends only
on the sum of the states of the cells in its neighborhood at the previous
time. Of the $2^{6}=64$ totalistic rules, 36 are independent. To
avoid confusion, we add the letter {}``\textit{T}'' to the Wolfram
code of the 5TCA rules, e.g., rule \textit{T}20.

\section{Derivation of networks from CA rules}

We consider a one-dimensional grid with \textit{N} cells, where each
cell is connected to its \textit{r} nearest neighbors with periodic
boundary conditions. Each cell state $x\in\left\{ 0,1\right\} $ evolves
by an identical CA rule function $f_{R}$, where \textit{R} denotes
its Wolfram code. After \textit{t} time steps, the configuration of
cells obtained from an initial one $\boldsymbol{\varphi\equiv x}(0)$
is given by $\boldsymbol{x}(t,\boldsymbol{\varphi})=\boldsymbol{f}_{R}^{t}(\boldsymbol{\varphi})$.
If $\boldsymbol{\varphi}_{i}$ denotes the initial configuration with
a changed state for cell i\textit{,} the difference between the configurations
after \textit{t} time steps from $\boldsymbol{\varphi}_{i}$ and $\boldsymbol{\varphi}$
can be written as

\begin{align}
\Delta_{i}\boldsymbol{x}(t,\boldsymbol{\varphi}) & \equiv\boldsymbol{x}(t,\boldsymbol{\varphi}_{i})+\boldsymbol{x}(t,\boldsymbol{\varphi})\;\left(\mathrm{mod}\:2\right)=\boldsymbol{f}_{R}^{t}(\boldsymbol{\varphi}_{i})+\boldsymbol{f}_{R}^{t}(\boldsymbol{\varphi})\;(\mathrm{mod}\:2)\label{eq:diff-f}\end{align}

\begin{flushleft}
where $\Delta_{i}\boldsymbol{x}(t,\boldsymbol{\varphi})$ characterizes
the influence of cell \textit{i} on other cells in the grid after
\textit{t} time steps. This influence represents the flow of information
from cell \emph{i} through the network. We define a matrix as \begin{eqnarray}
A_{R}(t,\boldsymbol{\varphi}) & \equiv & \left[\Delta_{0}\boldsymbol{x}(t,\boldsymbol{\varphi}),\Delta_{1}\boldsymbol{x}(t,\boldsymbol{\varphi}),\ldots,\Delta_{N-1}\boldsymbol{x}(t,\boldsymbol{\varphi})\right]^{T}\label{eq:adja}\\
 & = & \left[\Delta_{0}\boldsymbol{f}_{R}^{t}(\boldsymbol{\varphi}),\Delta_{1}\boldsymbol{f}_{R}^{t}(\boldsymbol{\varphi}),\ldots,\Delta_{N-1}\boldsymbol{f}_{R}^{t}(\boldsymbol{\varphi})\right]^{T},\label{eq:adja-f}\end{eqnarray}
where the transpose operation \textit{T} does not apply to the individual
elements $\Delta_{i}\boldsymbol{x}(t,\boldsymbol{\varphi})$. We treat
$A_{R}(t,\boldsymbol{\varphi})$ as the \textit{adjacency matrix}
of a network derived from the CA rule \textit{R}; an entry $a_{ij}=1$
if a directed edge from node \textit{i} to node \textit{j} exists,
and 0 otherwise. Because $\Delta_{i}\boldsymbol{x}(t,\boldsymbol{\varphi})$
depends on the initial configuration, $A_{R}(t,\boldsymbol{\varphi})$
gives a different network for each initial configuration. Although
this derived network is not identical with an analogous network derived
by considering the change of two or more cells, it captures essential
properties of its cellular automaton, as shown below. Some graphs
derived from ECA and 5TCA rules are presented in Fig.\ref{fig:ECA-g}
and Fig.\ref{fig:5TCA-g}, respectively. %
\begin{figure}
\begin{centering}
\subfloat[rule 62]{\includegraphics[width=2.5cm]{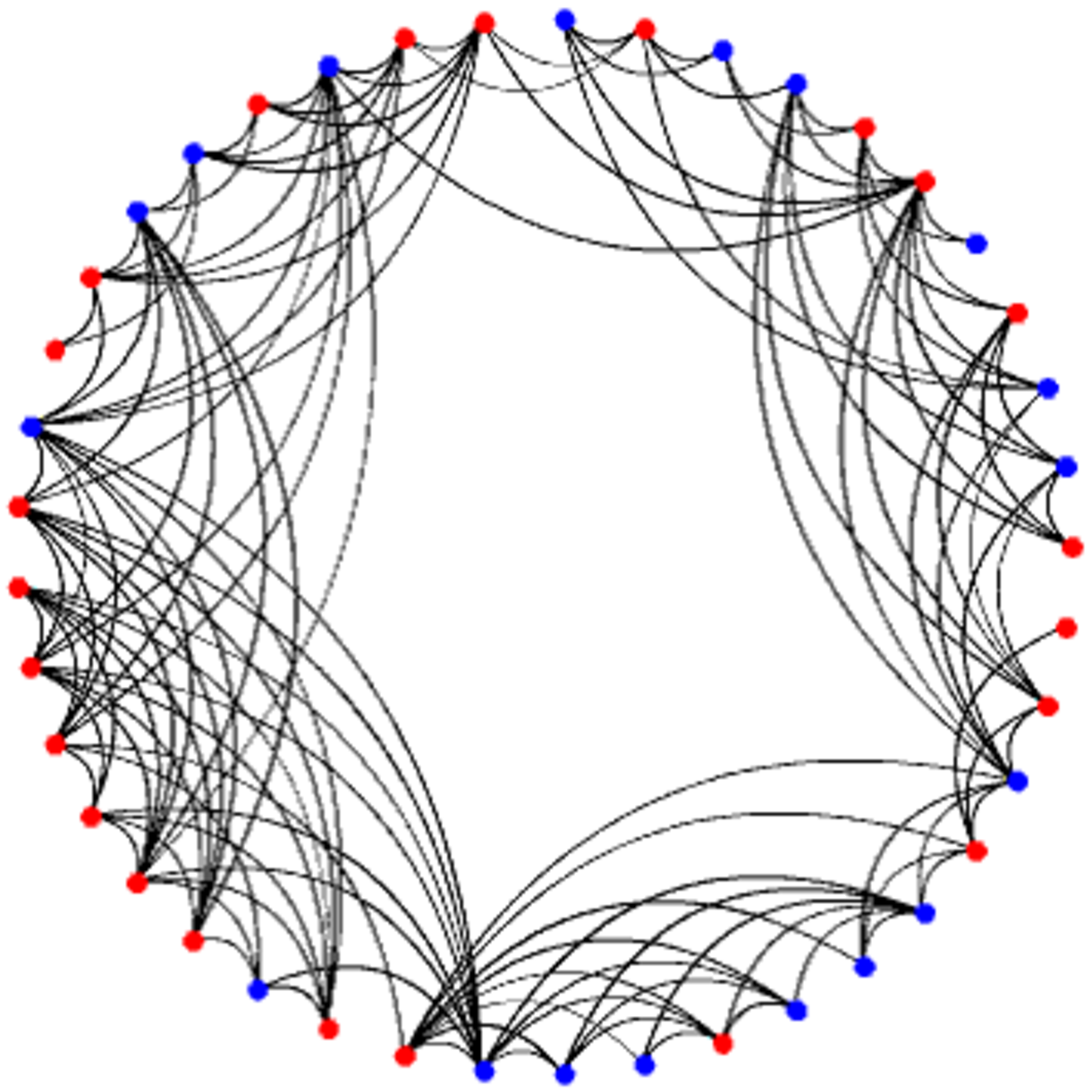}

}\subfloat[rule 30]{\includegraphics[width=2.5cm]{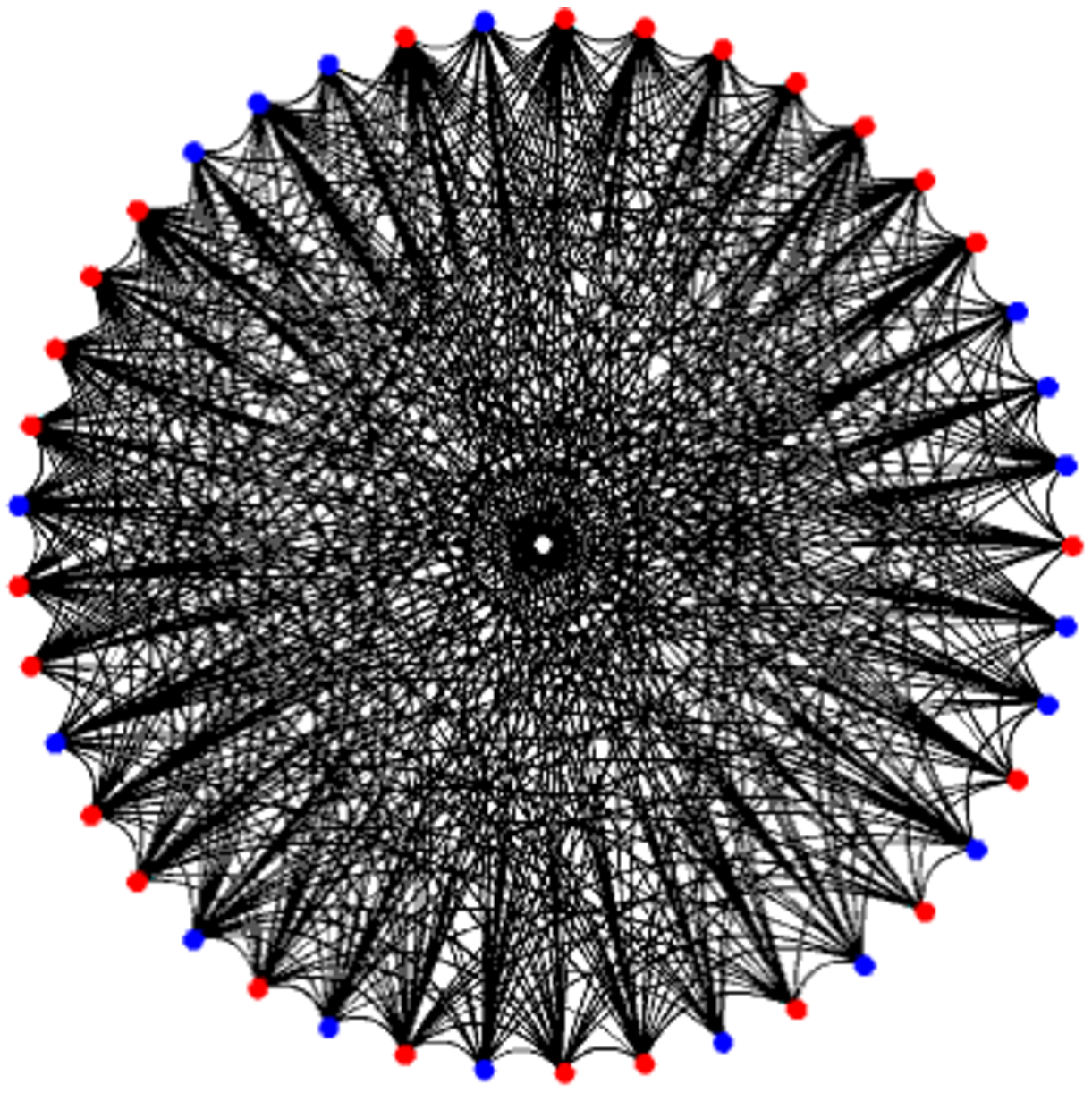}

}\subfloat[rule 90]{\includegraphics[width=2.5cm]{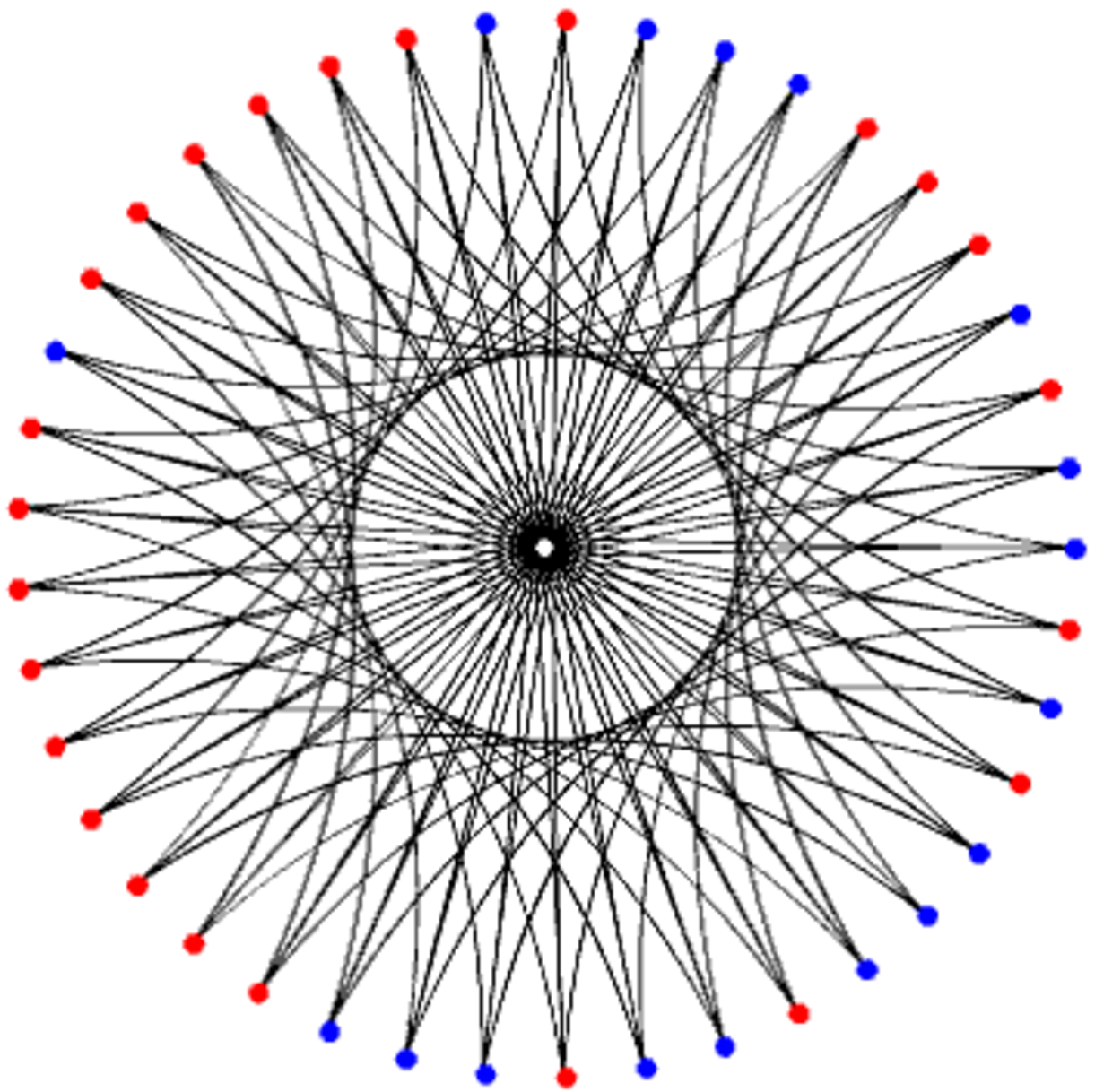}

}\subfloat[rule 54]{\includegraphics[width=2.5cm]{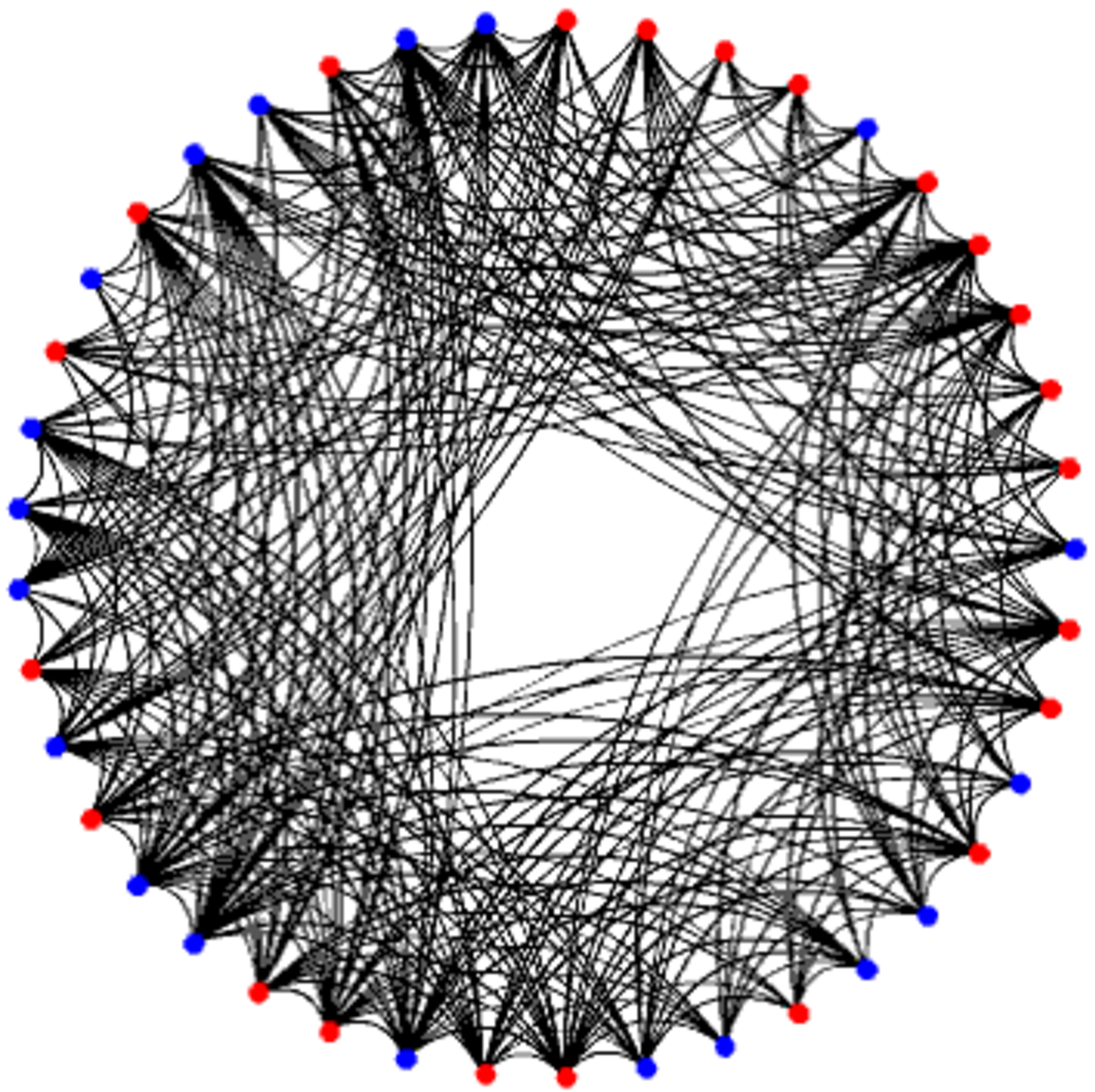}

}\subfloat[rule 110]{\includegraphics[width=2.5cm]{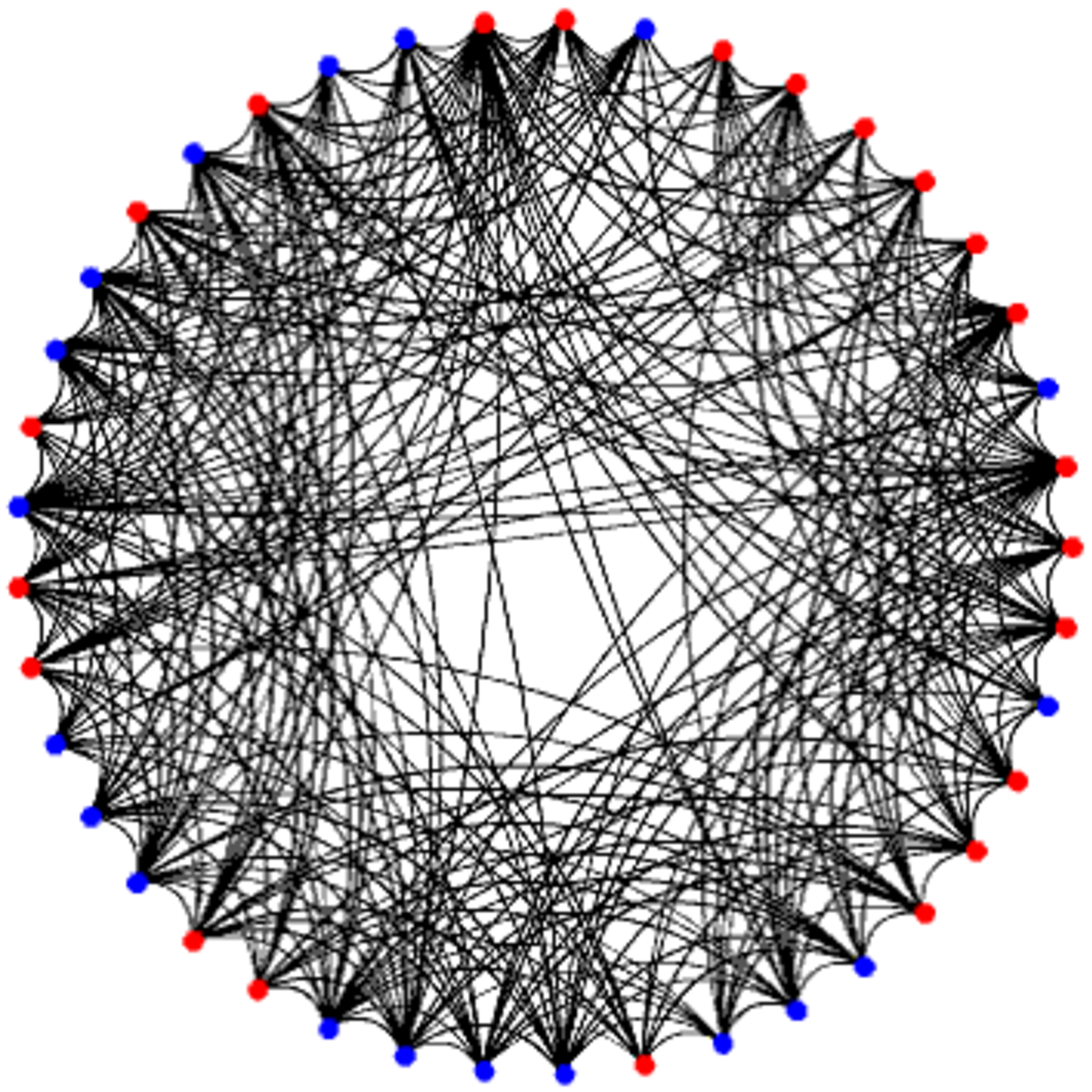}

}
\par\end{centering}

\caption{Examples of network graphs derived from ECA rules with $N=41$ and
$t=20$. The dots on the circumference of each graph correspond to
the nodes of the derived network, which in turn correspond to the
cells of the cellular automaton. The time \emph{t} is selected to
give each cell causal relationships with all other cells and to avoid
repetitions. \label{fig:ECA-g}}

\end{figure}
\begin{figure}
\begin{centering}
\subfloat[rule \emph{T}40]{\includegraphics[width=2.5cm]{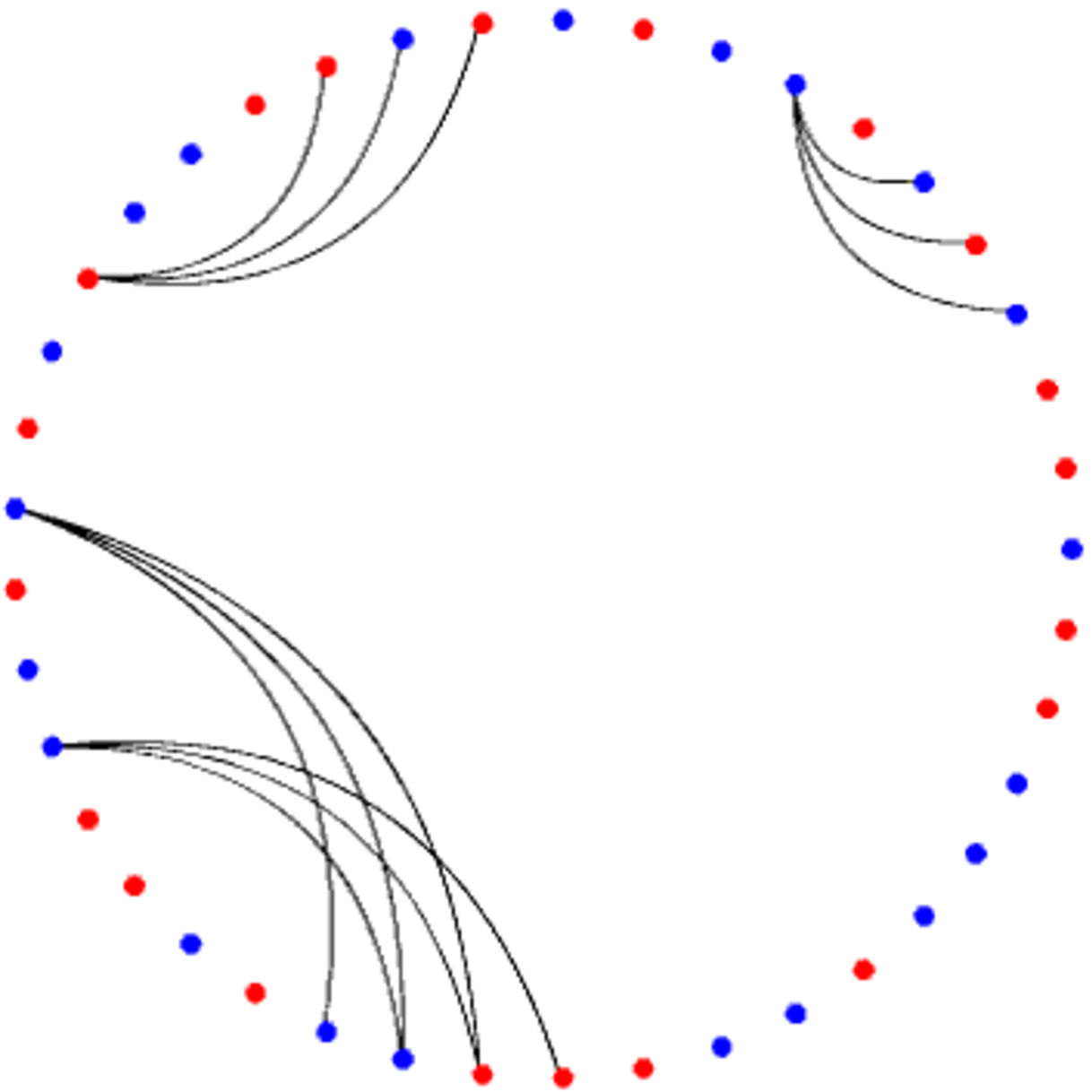}

}\subfloat[rule \emph{T}10]{\includegraphics[width=2.5cm]{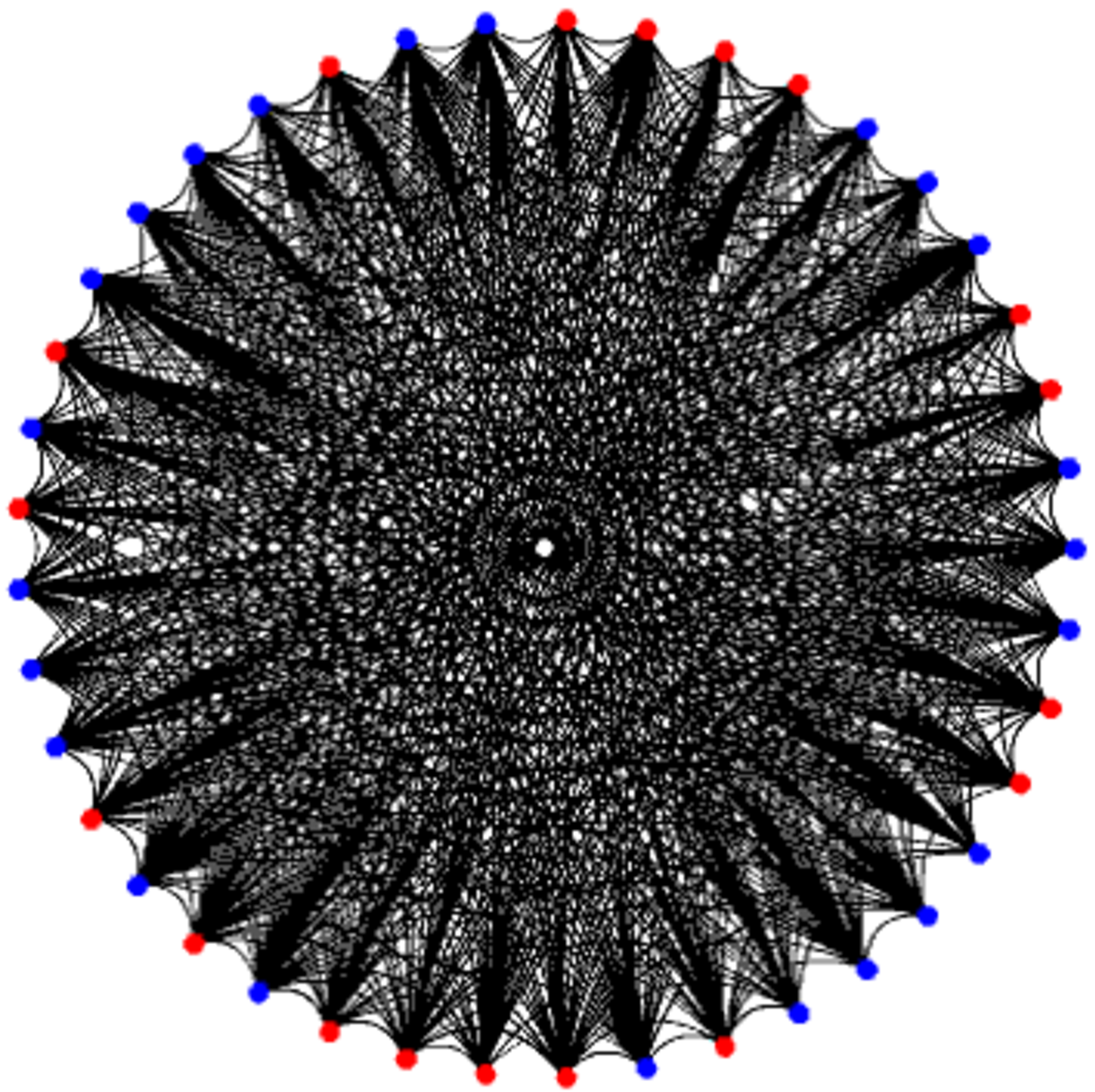}

}\subfloat[rule \emph{T}42]{\includegraphics[width=2.5cm]{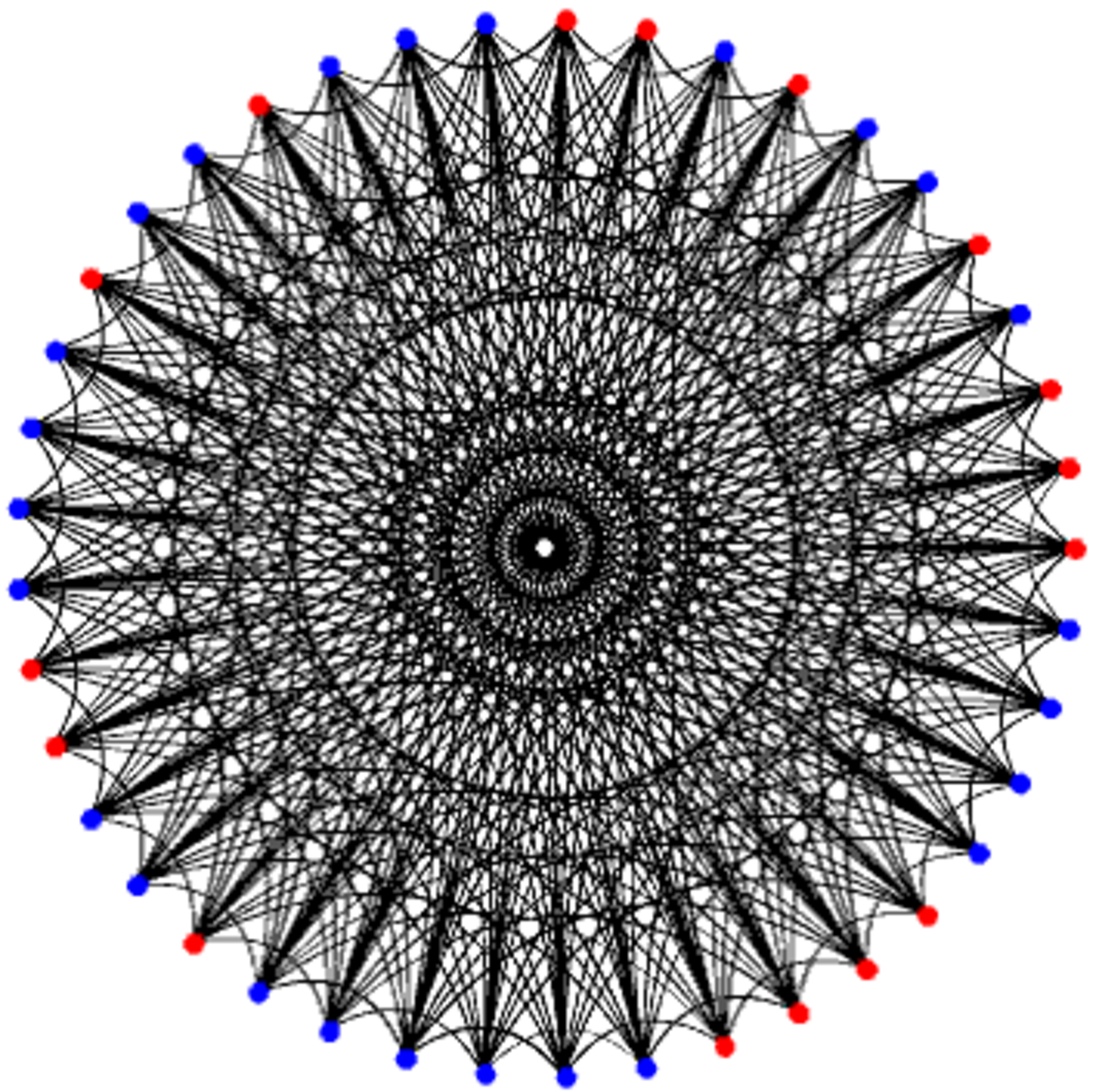}

}\subfloat[rule \emph{T}20]{\includegraphics[width=2.5cm]{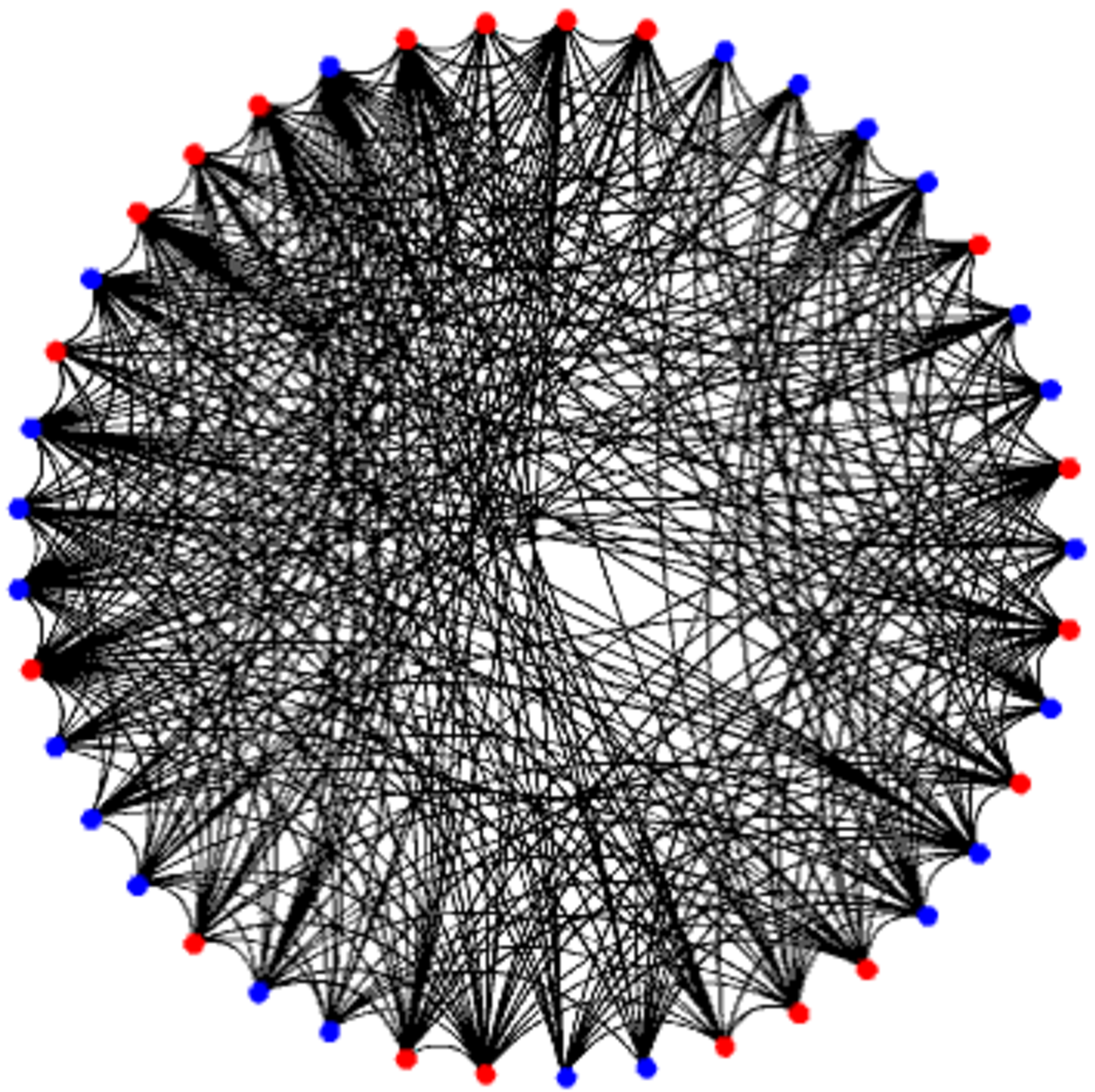}

}\subfloat[rule \emph{T}52]{\includegraphics[width=2.5cm]{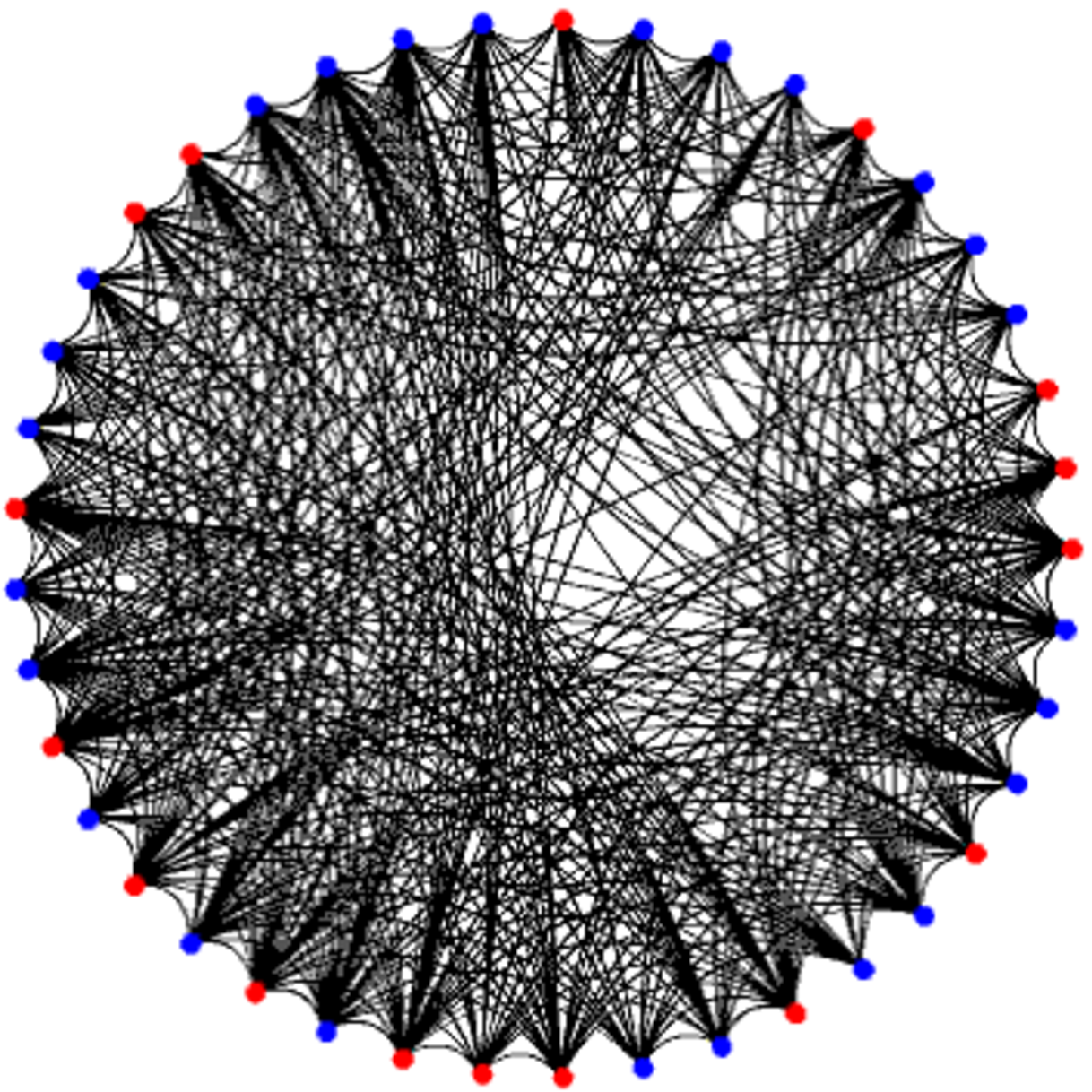}

}
\par\end{centering}

\caption{Examples of network graphs derived from 5TCA rules with $N=41$ and
$t=10$. \label{fig:5TCA-g}}

\end{figure}

\par\end{flushleft}

From Eqs.\eqref{eq:LR-t-f}, \eqref{eq:diff-f} and \eqref{eq:adja-f},
the adjacency matrix is transformed by the mirror operation as follows:

\begin{eqnarray}
\tilde{A}_{R}(t,\boldsymbol{\varphi}) & \equiv & \left[\Delta_{0}\tilde{\boldsymbol{f}}_{R}^{t}(\boldsymbol{\varphi}),\Delta_{1}\tilde{\boldsymbol{f}}_{R}^{t}(\boldsymbol{\varphi}),\ldots,\Delta_{N-1}\tilde{\boldsymbol{f}}_{R}^{t}(\boldsymbol{\varphi})\right]^{T}\nonumber \\
 & = & \left[\Delta_{N-1}\widetilde{\boldsymbol{f}_{R}^{t}(\tilde{\boldsymbol{\varphi}})},\Delta_{N-2}\widetilde{\boldsymbol{f}_{R}^{t}(\tilde{\boldsymbol{\varphi}})},\ldots,\Delta_{0}\widetilde{\boldsymbol{f}_{R}^{t}(\tilde{\boldsymbol{\varphi}})}\right]^{T}\label{eq:adja-LR-f}\\
 & = & \widetilde{A_{R}(t,\tilde{\boldsymbol{\varphi}})},\label{eq:adja-LR}\end{eqnarray}

\begin{flushleft}
where the subscript of $\Delta$ in Eq.\eqref{eq:adja-LR-f} labels
the state-changed elements of $\tilde{\boldsymbol{\varphi}}$ and
the $\sim$ operation acting on the adjacency matrix in the right-hand
side of Eq.\eqref{eq:adja-LR} represents the mirroring of elements
about both the horizontal and vertical axes. Similarly, the complement
of the adjacency matrix is obtained from Eqs.\eqref{eq:01-t-f}, \eqref{eq:diff-f}
and \eqref{eq:adja-f} as follows:
\par\end{flushleft}

\begin{eqnarray}
\bar{A}_{R}(t,\boldsymbol{\varphi}) & \equiv & \left[\Delta_{0}\bar{\boldsymbol{f}}_{R}^{t}(\boldsymbol{\varphi}),\Delta_{1}\bar{\boldsymbol{f}}_{R}^{t}(\boldsymbol{\varphi}),\ldots,\Delta_{N-1}\bar{\boldsymbol{f}}_{R}^{t}(\boldsymbol{\varphi})\right]^{T}\nonumber \\
 & = & \left[\Delta_{0}\overline{\boldsymbol{f}_{R}^{t}(\bar{\boldsymbol{\varphi}})},\Delta_{1}\overline{\boldsymbol{f}_{R}^{t}(\bar{\boldsymbol{\varphi}})},\ldots,\Delta_{N-1}\overline{\boldsymbol{f}_{R}^{t}(\bar{\boldsymbol{\varphi}})}\right]^{T}\nonumber \\
 & = & \left[\Delta_{0}\boldsymbol{f}_{R}^{t}(\bar{\boldsymbol{\varphi}}),\Delta_{1}\boldsymbol{f}_{R}^{t}(\bar{\boldsymbol{\varphi}}),\ldots,\Delta_{N-1}\boldsymbol{f}_{R}^{t}(\bar{\boldsymbol{\varphi}})\right]^{T}\label{eq:adja-01-f}\\
 & = & A_{R}(t,\bar{\boldsymbol{\varphi}}),\label{eq:adja-01}\end{eqnarray}

\begin{flushleft}
where Eq.\eqref{eq:adja-01-f} is derived from the invariance of the
$modulo$ operation in Eq.\eqref{eq:diff-f} under complementation.
In general, the graph of the complement $\bar{A}_{R}(t,\boldsymbol{\varphi})$
is not identical with the {}``graph complement'' \citep{S.Pemmaraju2003}
of $A_{R}(t,\boldsymbol{\varphi})$. Finally, the mirror-complement
of the adjacency matrix is
\par\end{flushleft}

\begin{eqnarray}
\bar{\tilde{A}}_{R}(t,\boldsymbol{\varphi}) & = & \widetilde{A_{R}(t,\bar{\tilde{\boldsymbol{\varphi}}})}=\tilde{\bar{A}}_{R}(t,\boldsymbol{\varphi}).\label{eq:adja-LR-01}\end{eqnarray}

An important property of the adjacency matrix may be derived for self-complementary
rule functions. Adding the transformations \eqref{eq:LR}-\eqref{eq:LR-01},
we define the \textit{diminished-radix complement} of the rule function
$f_{R}$ as

\begin{equation}
\hat{f_{R}}(x_{i-r},...,x_{i},...,x_{i+r})\equiv f_{R}(\bar{x}_{i-r},...,\bar{x}_{i},...,\bar{x}_{i+r}).\end{equation}

\begin{flushleft}
If a mapping $\boldsymbol{f}_{R}$ is \emph{self-complement}ary, i.e.
$\boldsymbol{f}_{R}=\bar{\boldsymbol{f}}_{R}$, the mapping $\hat{\boldsymbol{f}_{R}}^{t}$
obtained from $\hat{f_{R}}$ satisfies the equation
\par\end{flushleft}

\begin{eqnarray}
\hat{\boldsymbol{f}}_{R}^{t}(\boldsymbol{\varphi}) & = & \begin{cases}
\overline{\boldsymbol{f}_{R}^{t}(\boldsymbol{\varphi})} & \;\textrm{for odd }t\\
\boldsymbol{f}_{R}^{t}(\boldsymbol{\varphi}) & \;\textrm{for even }t.\end{cases}\label{eq:even-odd-drc-f}\end{eqnarray}

\begin{flushleft}
Therefore, $\Delta_{i}\hat{\boldsymbol{f}}_{R}^{t}(\boldsymbol{\varphi})$
is equal to $\Delta_{i}\boldsymbol{f}_{R}^{t}(\boldsymbol{\varphi})$
for all \textit{t}, and the adjacency matrix $\hat{A}_{R}(t,\boldsymbol{\varphi})$
defined by $\hat{\boldsymbol{f}}_{R}^{t}(\boldsymbol{\varphi})$ is
identical with $A_{R}(t,\boldsymbol{\varphi})$. Although this property
of the adjacency matrix makes the rule $\hat{f_{R}}$ indistinguishable
from rule \textit{R}, this degeneracy is not a drawback of our method
but rather a new way of detecting similarity among CA rules, as described
below.
\par\end{flushleft}

Another interesting property of our approach pertains to \emph{additive
mappings} which satisfy the equation $\boldsymbol{f}_{R}(\boldsymbol{x}+\boldsymbol{y}\;(\mathrm{mod}\:2))=\boldsymbol{f}_{R}(\boldsymbol{x})+\boldsymbol{f}_{R}(\boldsymbol{y})\;(\mathrm{mod}\:2)$.
Eq.\eqref{eq:diff-f} leads to the result $\Delta_{i}\boldsymbol{x}(t,\boldsymbol{\varphi})=\boldsymbol{f}_{R}^{t}(\boldsymbol{0}_{i})$,
where $\boldsymbol{0}_{i}$ is the \textit{null} configuration except
at cell \emph{i}. Thus, the adjacency matrix \eqref{eq:adja-f} is
independent of the initial configuration and all nodes of the derived
network are equivalent. Hence, the same number of edges connects to
each node. If an edge from node \textit{i} to node \textit{j} exists,
an edge also exists in the opposite direction; the network is undirected,
which means that the adjacency matrix is symmetric. For example, rule
90 is additive and, for this reason, corresponds to the geometrical
graph shown in Fig.1(c). Furthermore, if cell\emph{ i} and cell\emph{
j} change their states in the initial configuration $\boldsymbol{\varphi}$,
the difference between the configurations $\boldsymbol{\varphi}_{ij}$
and $\boldsymbol{\varphi}$ after\emph{ t} time steps satisfies the
equation

\begin{eqnarray}
\Delta_{ij}\boldsymbol{x}(t,\boldsymbol{\varphi}) & = & \boldsymbol{f}_{R}^{t}(\boldsymbol{\varphi}_{ij})+\boldsymbol{f}_{R}^{t}(\boldsymbol{\varphi})\;(\mathrm{mod}\:2)=\boldsymbol{f}_{R}^{t}(\boldsymbol{0}_{ij})=\boldsymbol{f}_{R}^{t}(\boldsymbol{0}_{i})+\boldsymbol{f}_{R}^{t}(\boldsymbol{0}_{j})\;(\mathrm{mod}\:2)\nonumber \\
 & = & \Delta_{i}\boldsymbol{x}(t,\boldsymbol{0})+\Delta_{j}\boldsymbol{x}(t,\boldsymbol{0})\;(\mathrm{mod}\:2)\\
 & = & (A_{R}(t))_{i}+(A_{R}(t))_{j}\;(\mathrm{mod}\:2).\end{eqnarray}

\begin{flushleft}
where $(A)_{i}$ denotes \textit{i}-th row vector of matrix $A$.
Consequently, the adjacency matrix for additive mappings can represent
the influence of changes at more than two cells.
\par\end{flushleft}

\section{Properties of derived networks}

Following the Wolfram's classification, we examine some typical rules
of ECA and 5TCA. There is no reason to restrict our discussion to
even-numbered rules. We use two basic parameters, the efficiency and
the degree distribution, to investigate the properties and complexities
of derived networks.

A descriptive network parameter is \textit{\emph{the distribution}}
of degrees. $P_{R}(k_{in})$ and $P_{R}(k_{out})$ indicate the probabilities
of a node having an in-degree $k_{in}$ and an out-degree $k_{out}$,
respectively. In particular, scale-free networks are the class of
networks whose degree distribution is a power-law: $P(k)\sim k^{-\gamma}$,
where $\gamma$ is called the \emph{scale-free exponent}. Small-world
networks can be categorized by their average shortest path length
$l=<d_{ij}>$, where $d_{ij}$ is the length of the shortest path
between node \textit{i} and node \textit{j}. To sidestep any divergence
of the $d_{ij}$, we consider their harmonic mean, which we use to
define the network efficiency $E=\frac{1}{N(N-1)}\sum_{i\neq j}\frac{1}{d_{ij}}$
\citep{Latora2001,S.Boccaletti2006}.

Because these parameters are calculated for adjacency matrices obtained
from randomly generated initial configurations, the transformed matrices
defined in Eqs.\eqref{eq:adja-LR}, \eqref{eq:adja-01} and \eqref{eq:adja-LR-01}
give assuredly equivalent results with the original adjacency matrix.
Hence, the dependence of efficiency on grid size \textit{N} is very
useful for the classification of CA rules. Further insight is obtained
by using the degree distribution to characterize the CA rules; for
example, random and scale-free distributions are found in networks
derived from chaotic and complex rules, respectively.

Pairs of the diminished-radix complement of self-complementary rules
of ECA and 5TCA are listed in Table 1. %
\begin{table}
\begin{centering}
\subfloat[ECA pairs]{\begin{centering}
\begin{tabular}{|>{\centering}p{0.6in}|>{\centering}p{0.6in}|}
\hline 
15(85) & 240(170)\tabularnewline
\hline 
23 & 232\tabularnewline
\hline 
43(113) & 212(142)\tabularnewline
\hline 
51 & 204\tabularnewline
\hline 
77 & 178\tabularnewline
\hline 
105 & 150\tabularnewline
\hline
\end{tabular}
\par\end{centering}

}\subfloat[5TCA pairs]{\begin{centering}
\begin{tabular}{|>{\centering}p{0.6in}|>{\centering}p{0.6in}|}
\hline 
7 & 56\tabularnewline
\hline 
11 & 52\tabularnewline
\hline 
21 & 42\tabularnewline
\hline 
25 & 38\tabularnewline
\hline 
\multicolumn{1}{>{\centering}p{0.6in}}{} & \multicolumn{1}{>{\centering}p{0.6in}}{}\tabularnewline
\multicolumn{1}{>{\centering}p{0.6in}}{} & \multicolumn{1}{>{\centering}p{0.6in}}{}\tabularnewline
\end{tabular}
\par\end{centering}

}
\par\end{centering}

\caption{The diminished-radix complements pairs of self-complement rules of
ECA and 5TCA. The rule inside the parentheses is mirror equivalent
to the one outside the parentheses. }

\end{table}
 Although the individual rules of a pair have different CA patterns,
they have similar statistical properties and belong to the same Wolfram
class. As indicated by Eq.\eqref{eq:even-odd-drc-f}, this correspondence
is obvious from the invariance of the time evolution under the exchange
of zero and one (complementation).

\subsection*{Networks derived from ECA and 5TCA rules}

As shown in Figs.\ref{fig:ECA-g} and \ref{fig:5TCA-g}, the topology
of the derived network reflects properties of the corresponding CA
rule. %
\begin{figure}
\begin{centering}
\includegraphics[angle=-90,width=9cm]{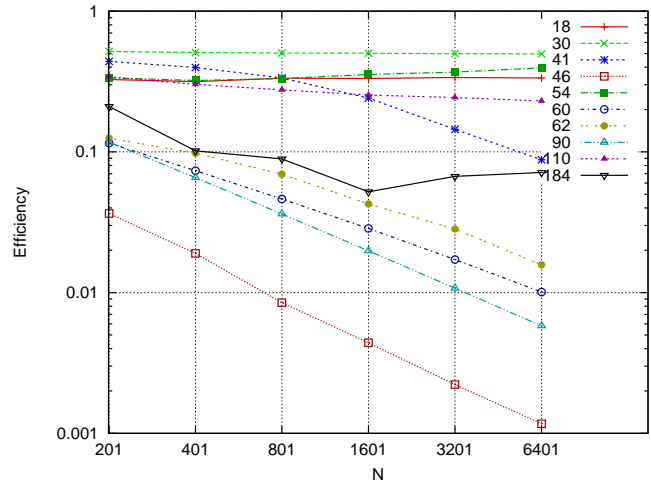}
\par\end{centering}

\caption{Efficiencies of networks derived from ECA rules, representing the
average of ten sampled networks. \label{fig:eca-eff-all}}

\end{figure}
\begin{figure}
\begin{centering}
\includegraphics[angle=-90,width=9cm]{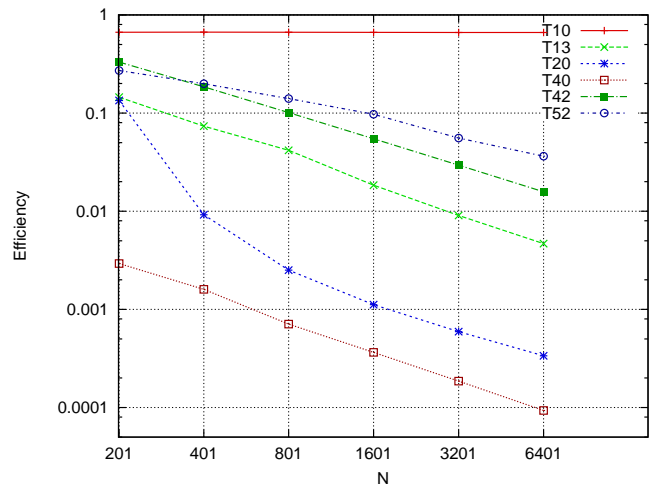}
\par\end{centering}

\caption{Efficiencies of networks derived from 5TCA rules, representing the
average of ten sampled networks. \label{fig:5tca-eff-all}}

\end{figure}
 Fig.\ref{fig:eca-eff-all} and Fig.\ref{fig:5tca-eff-all} plots
the efficiencies of illustrative ECA and 5TCA rules, respectively.
The efficiency values can be categorized into several types. All class
I CA have zero efficiency, whereas class III CA are characterized
by a high, \emph{N}-independent efficiency (i.e. short average path
length), except for rule 90, 60 and \emph{T}42. For class II CA, the
efficiency decreases as $N^{-1}$, with the exception of rule 184.
This \emph{N}-dependence of class II efficiency can be explained intuitively
as follows. Because nontrivial patterns of class II CA are localized,
the derived networks are disconnected; hence, the number of edges
is proportional to $N$, whereas the total number of node pairs is
proportional to $N^{2}$. Therefore, the contribution of edges to
the efficiency decreases as $N^{-1}$. Although it is somewhat difficult
to decide empirically how to classify rule 41, the efficiency values
above \textit{$N>1600$} provide evidence that it belongs to class
II. 

The efficiency of rule 110 and \emph{T}52, which is thought to belong
to class IV, varies as $N^{-0.117}$ and $N^{-0.53}$(as estimated
by least-squares fitting), respectively; these \emph{N}-dependence
indicates that they lies on the {}``edge of chaos'' \citep{Langton1990}.
In contrast, rule 54 appears to be indistinguishable from class III
rules. Furthermore, the efficiency of rule \textsl{T}20 is almost
proportional to $N^{-1}$ for \emph{N}>1600. Thus it is difficult
to distinguish rule \emph{T}20 from class II rules using only its
efficiency dependence. 

Rules 90, 60 and \emph{T}42 are exceptional. Because they are additive,
each node has the same number of edges, and the total number of edges
is proportional to $N$. Hence, their efficiencies have similar tendencies
to those of class II CA. However, in the case of $t=2^{n}-1$ (where
\textit{n} is a positive integer) and $N=2t+1$, each node of rule
90 has \emph{$2^{n}=(N-1)/2$} edges, which means that its numbers
of total edges are proportional to $N^{2}$. Consequently, the efficiency
of rule 90 oscillate between class II and class III regions. Similar
oscillations of the efficiencies of rule 60 and \emph{T}42 are confirmed.
Another exception is rule 184, which is usually called the \textquotedbl{}traffic
rule\textquotedbl{}; its derived network has random values of efficiency,
indicating strong dependence on the initial configuration. This random
behavior can be interpreted as critical phenomena of the phase transition
described in \citet{Wolfram2002}.

\begin{figure}
\begin{centering}
\subfloat[rule 62 with\emph{ $N=6401$} and $t=3200$]{\includegraphics[angle=-90,width=6cm]{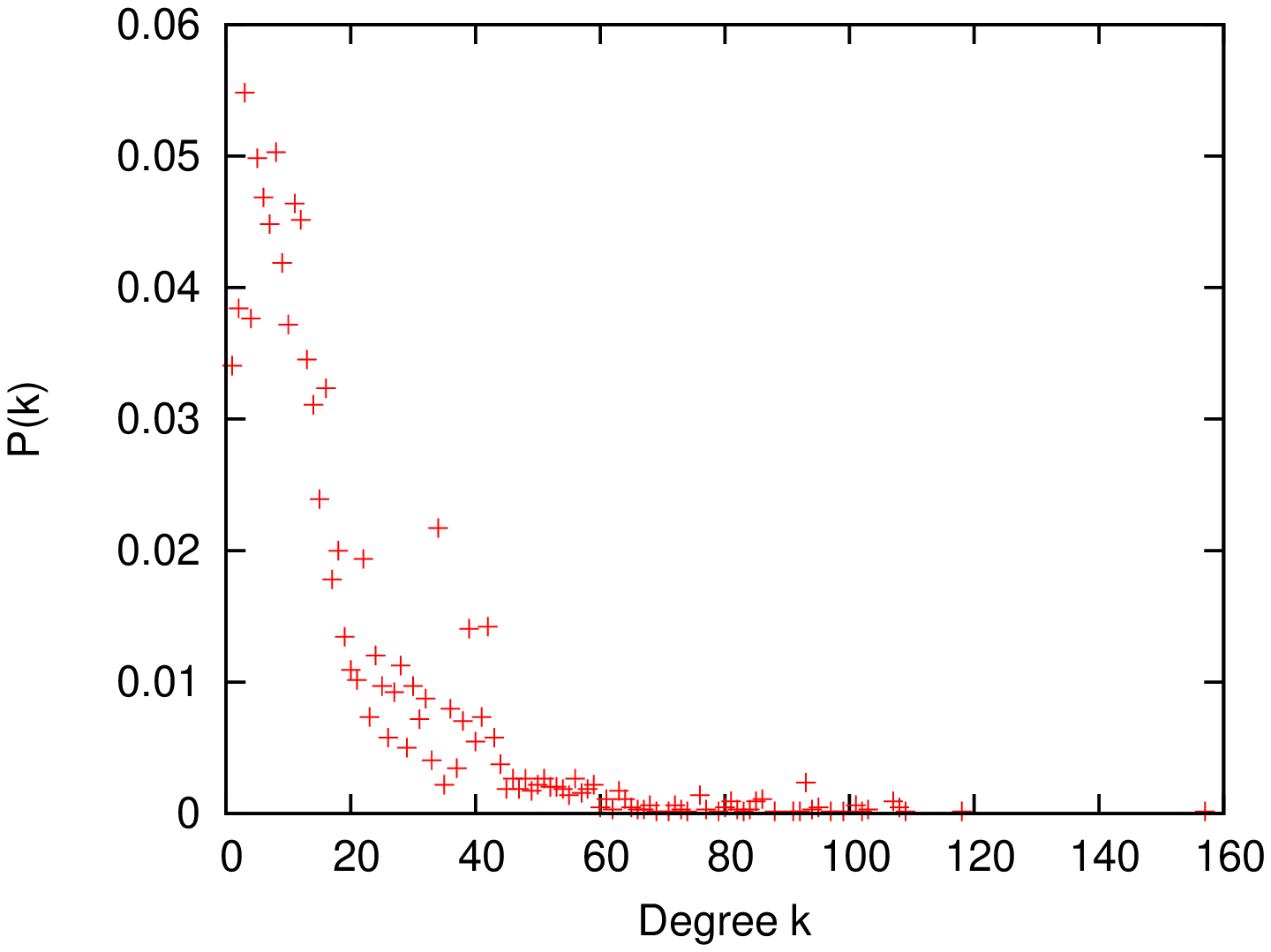}

}\subfloat[rule \emph{T}13 with \emph{$N=6401$} and $t=1600$]{\includegraphics[angle=-90,width=6cm]{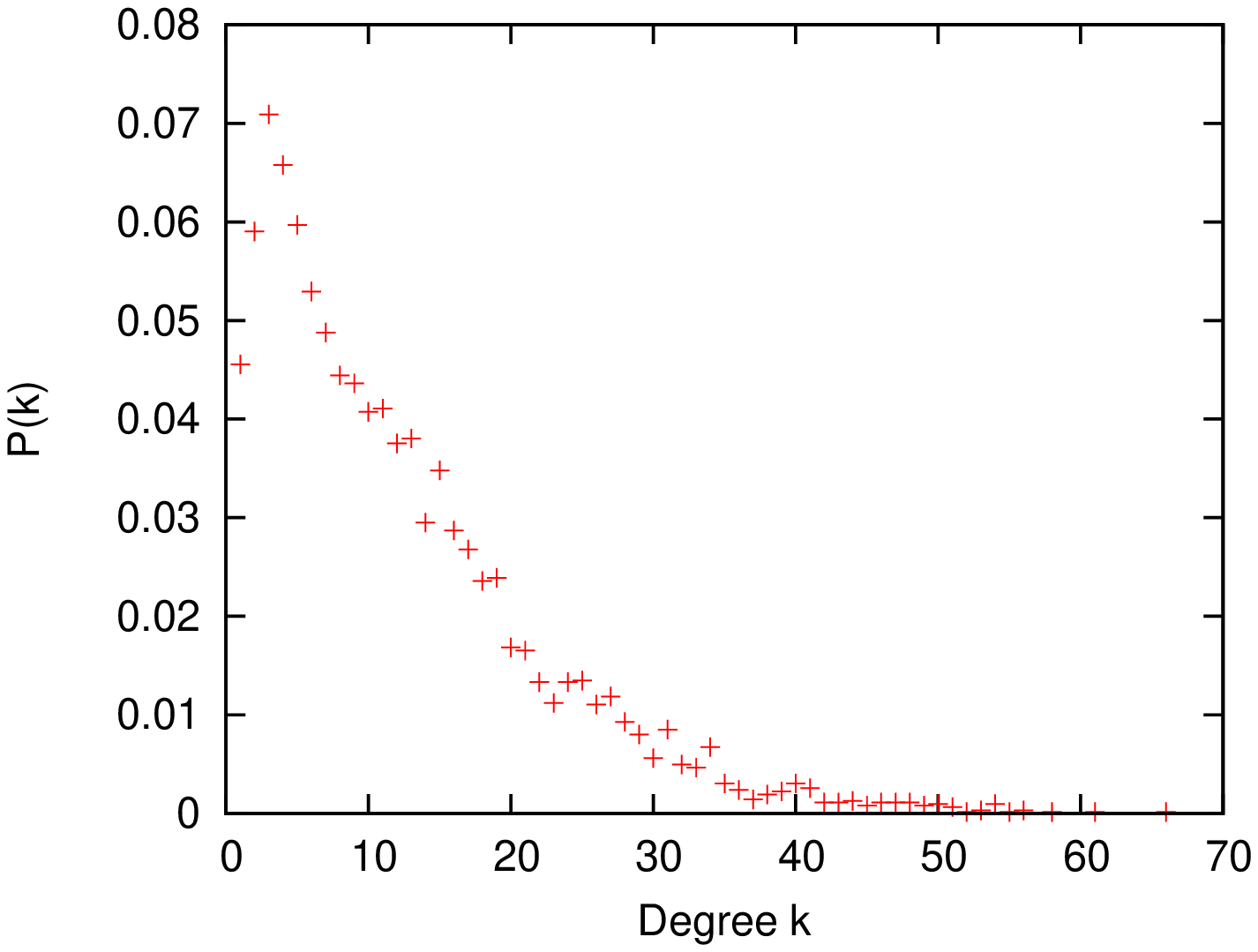}

}
\par\end{centering}

\caption{In-degree distributions of networks derived from ECA and 5TCA class
II rules.\label{fig:in-dd-c2}}

\end{figure}

\begin{figure}
\begin{centering}
\subfloat[rule 30 with \emph{$N=6401$} and $t=3200$]{\includegraphics[angle=-90,width=6cm]{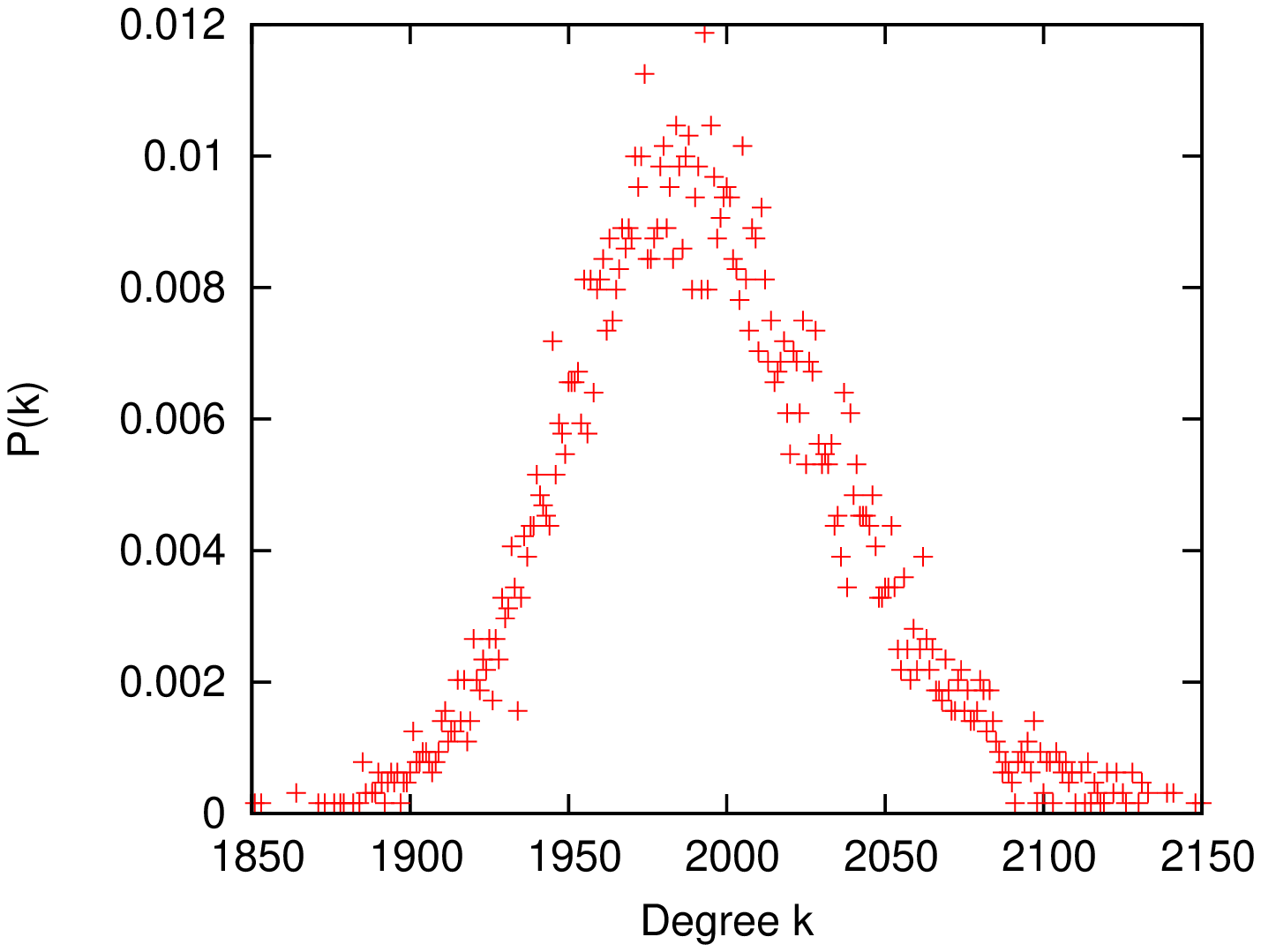}}\subfloat[rule \emph{T}10 with \emph{$N=6401$} and $t=1600$]{\includegraphics[angle=-90,width=6cm]{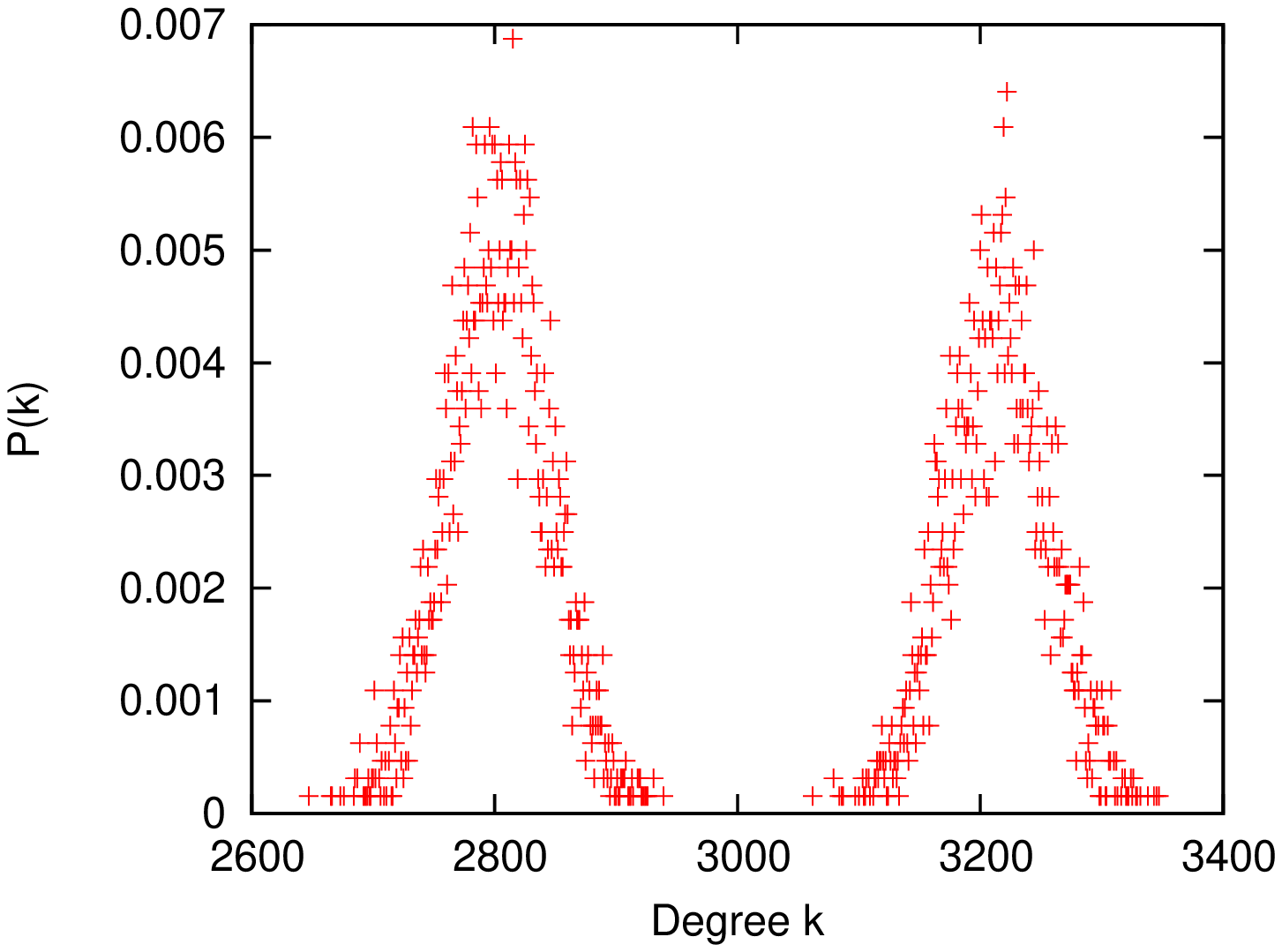}

}
\par\end{centering}

\caption{In-degree distributions of networks derived from ECA and 5TCA class
III rules.\label{fig:in-dd-c3}}

\end{figure}

The in-degree distributions of rule 62 and \emph{T}13 (class II) and
those of rule 30 and \emph{T}10 (class III) are illustrated in Fig.\ref{fig:in-dd-c2}
and Fig.\ref{fig:in-dd-c3}, respectively. %
 Rule 62 is a class II rule but has a large components of class III
character \citep{Kayama1995}, which may account for its long-tail
distribution. Rule 30 is well-known for its random behavior, from
which a random network with a Poisson distribution of degrees is derived.
Similarly, rule \textit{T}10 has two peaks, each fitting a Poisson
distribution; this double-peak property corresponds to the double-triangle
structure of its CA patterns.

\begin{figure}
\begin{centering}
\subfloat[rule \emph{T}20]{\includegraphics[angle=-90,width=6cm]{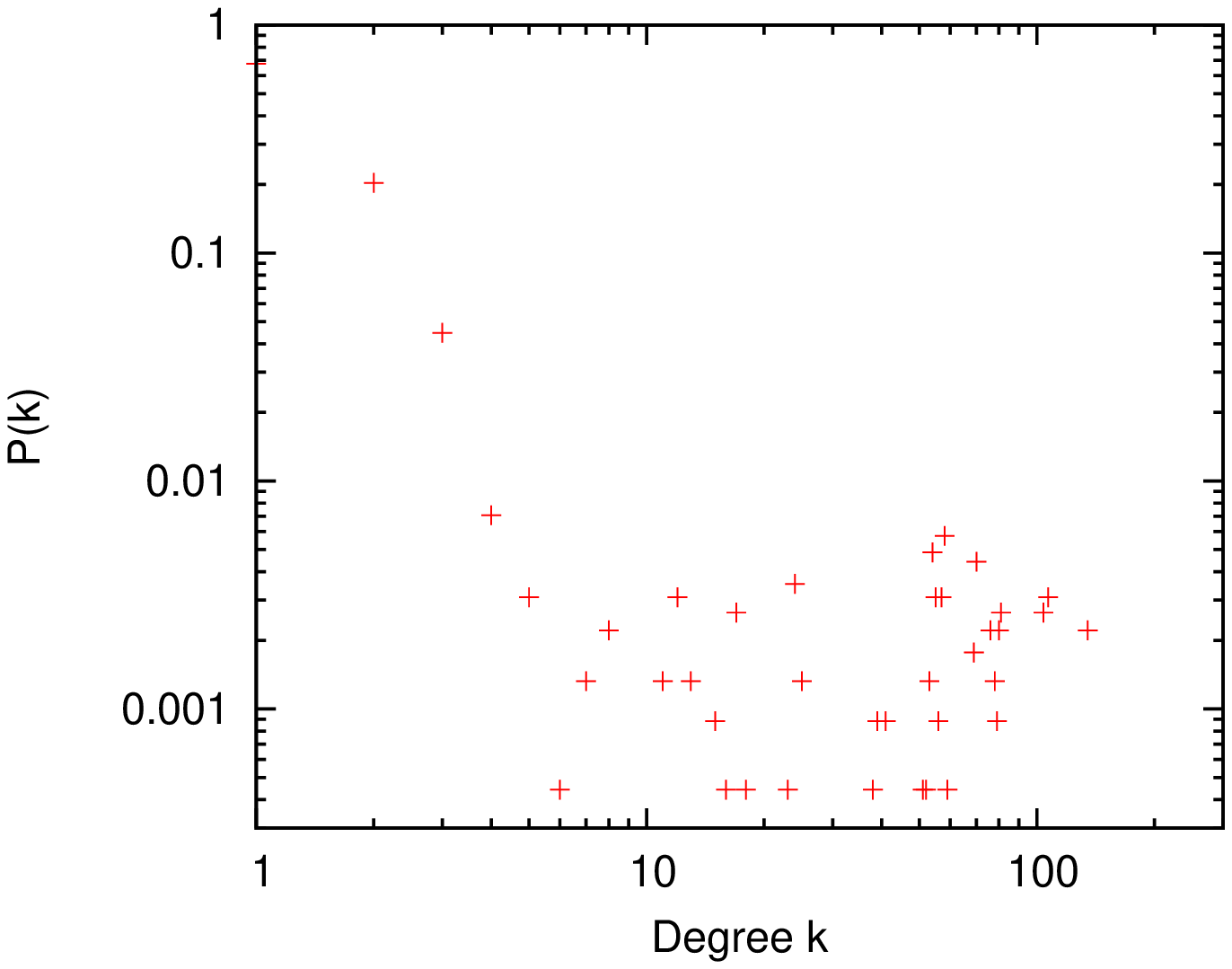}

}\subfloat[rule \emph{T}52]{\includegraphics[angle=-90,width=6cm]{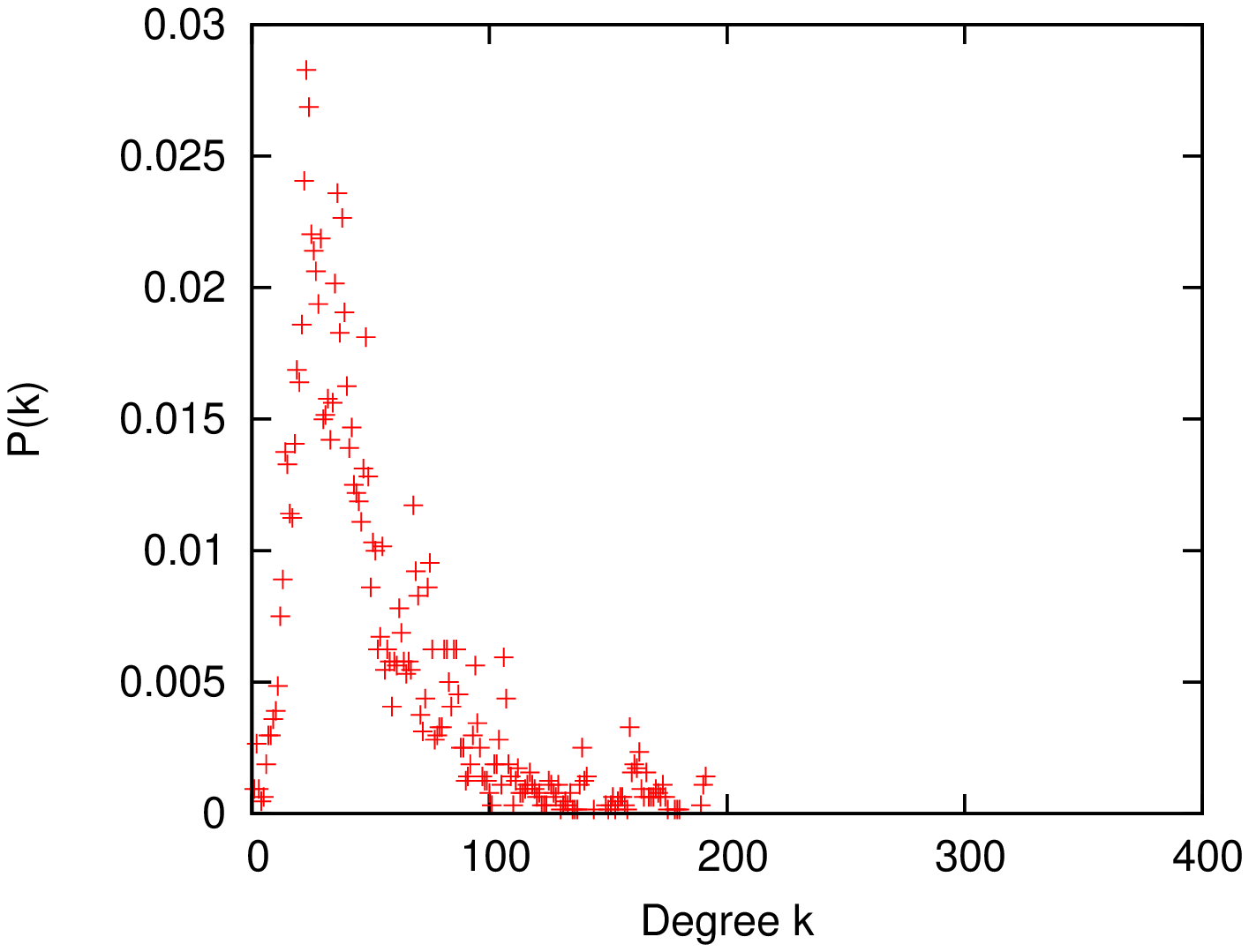}

}
\par\end{centering}

\caption{In-degree distributions of networks derived from 5TCA class IV rules
with \emph{$N=6401$} and $t=1600$. (a) The distribution of rule
\emph{T}20 plotted on log-log scale exhibits scale-free behavior.\label{fig:in-dd-c4}}

\end{figure}

\begin{figure}
\begin{centering}
\includegraphics[angle=-90,width=8cm]{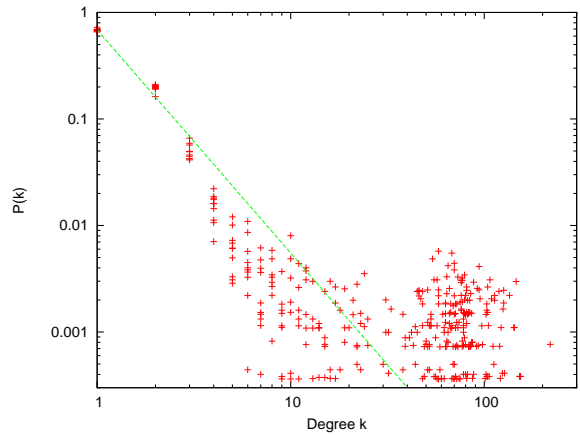}
\par\end{centering}

\caption{In-degree distribution of networks derived from rule \emph{T}20. The
results for ten sampled networks are illustrated with a best-fitting
power-law, yielding a scale-free exponent $\gamma=2.096$.\label{fig:in-dd-t20-SF}}

\end{figure}

Fig.\ref{fig:in-dd-c4} shows that class IV rules exhibit exponential
decays and long-tail distributions. %
 Fig.\ref{fig:in-dd-t20-SF} is an unaveraged distribution of ten
sampled networks of rule \textit{T}20, from which we obtain the scale-free
exponent $\gamma=2.096$, based on linear fitting. Hence, a scale-free
network can be derived from a class IV rule. %
 However, the $N^{-1}$ efficiency dependence of rule \emph{T}20 suggests
that the network derived from rule \emph{T}20 is disconnected in many
components, which contrasts with other scale-free networks such as
the BA model. We have confirmed that the network derived from rule
\textit{T}20 with $N=6401$ and $t=1600$ has dozens of disconnected
components. In other words, the network can be rendered scale-free
and connected by adding, at most, a few dozen edges. By examining
Figs.\ref{fig:in-dd-c3}(b), \ref{fig:in-dd-c4}(b) and \ref{fig:in-dd-c4}(a),
it is apparent that the two peaks gradually collapse; the left peak
becomes the scale-free distribution, whereas the right peak becomes
a structure at the lower slope. This change of network structure may
represent a phase transition from a chaotic phase to a periodic phase.

\section{Conclusions and discussion}

In this article, we have proposed a method for deriving networks from
binary CA rules. Networks of representative ECA and 5TCA rules have
been shown and we have discussed their properties in terms of two
parameters; their efficiency and their degree distribution. The efficiency
parameters appear to be useful in classifying CA rules. Representative
degree distributions of complex networks have also been determined,
exhibiting Poisson, long-tail and scale-free distributions. our approach
has other interesting aspects, such as a new way to characterize additive
rules and the similarity of CA rules. We conclude that our derived
network is an effective representation of CA.

A survey of all ECA and 5TCA rules will be reported in a full-length
paper. However, our method can be applied or extended to other CA
rules. For example, it is straightforward to apply it to binary CA
with two or more dimensions and many neighbors, such as the {}``Game
of Life''. The extension of the adjacency matrix to CA with three
or more states may yield weighted networks. 

One remaining problem is how one is to understand the appearance of
a scale-free degree distribution. The fundamental elements leading
to emergent scale-free properties are thought to be the growth of
the network and the preferential attachment of its links. The former
corresponds to the time evolution of CA and the latter may derive
from the CA rule and the randomly chosen initial configuration. However,
if the following correspondences between dynamical phases of complex
systems and types of complex network are assumed,
\begin{itemize}
\item Periodic phase $\Leftrightarrow$ disconnected and localized network
\item Chaotic phase $\Leftrightarrow$ random network
\item Critical phase $\Leftrightarrow$ scale-free network,
\end{itemize}
it may be possible to describe phase-transition phenomena as evolutions
and transformations of network structures. If so, the scale-free network
may be an intermediate structure of the description. In any event,
these correspondences and the appearance of the scale-free degree
distribution warrant further investigation.

\section*{Acknowledgments}

I wish to acknowledge valuable discussions with Yasumasa Imamura on
this topic.

%

\end{document}